%% file: main.tex
\documentclass[10pt,twocolumn,letterpaper]{article}

\usepackage[pagenumbers]{cvpr} 

\usepackage{graphicx}
\usepackage{amsmath}
\usepackage{amssymb}
\usepackage{booktabs}
\usepackage{amsthm}

\usepackage[pagebackref=true,breaklinks=true,letterpaper=true,colorlinks=true,bookmarks=false,linkcolor=crimson,citecolor=green]{hyperref}

\usepackage[capitalize]{cleveref}
\crefname{section}{Sec.}{Secs.}
\Crefname{section}{Section}{Sections}
\Crefname{table}{Table}{Tables}
\crefname{table}{Tab.}{Tabs.}

\usepackage[square,sort, numbers]{natbib}
\usepackage{times}
\usepackage{epsfig}
\usepackage{graphicx}
\usepackage{booktabs}
\usepackage{makecell}
\usepackage{multirow}
\usepackage{amsmath}
\usepackage{amssymb}
\usepackage[dvipsnames,table,xcdraw]{xcolor}
\usepackage{transparent}
\usepackage{mathtools}
\usepackage{pifont} 
\usepackage{subcaption}

\definecolor{crimson}{rgb}{0.86, 0.08, 0.24}
\definecolor{dodgerblue}{rgb}{0.12, 0.56, 1.0}
\definecolor{myblue}{rgb}{0.1, 0.3, 1.0}

\newcommand{\eccen}{\texttt{\textbf{Eccentric}}\@\xspace}
\newcommand{\Eccen}{\texttt{\textbf{Eccentric}}\@\xspace}
\newcommand{\ecc}{\texttt{\textbf{ECC}}~}
\newcommand{\eccs}{\texttt{\textbf{ECC}}'s~}
\newcommand{\ecci}{$\texttt{\textbf{ECC}}_\text{I}$~}
\newcommand{\ecca}{$\texttt{\textbf{ECC}}_\text{A}$~}
\newcommand{\eccd}{$\texttt{\textbf{ECC}}_\text{D}$~}
\input{commands}

\newlength\MAX  
\newcommand*\Chartr[1]{#1~\rlap{\textcolor{gray!30}{\rule{\MAX}{2ex}}}\transparent{0.6}{\textcolor{RubineRed}{\rule{#1\MAX}{2ex}}}}
\newcommand*\Chartrt[1]{\rlap{\textcolor{gray!30}{\rule{\MAX}{2ex}}}\transparent{0.6}{\textcolor{RubineRed}{\rule{#1\MAX}{2ex}}}}

\newcommand*\Chartgt[1]{\rlap{\textcolor{gray!30}{\rule{\MAX}{2ex}}}\transparent{0.6}{\textcolor{Green}{\rule{#1\MAX}{2ex}}}}

\begin{document}

\title{ECCENTRIC: Edge-Cloud Collaboration Framework for Distributed Inference Using Knowledge Adaptation}
\author{
  Mohammad Mahdi Kamani \quad Zhongwei Cheng \quad Lin Chen\\
  Wyze Labs, Inc.
}
\maketitle

\input{0-abstract}

\input{1-introduction}
\input{3-eccenteric}

\input{4-training}

\input{5-eval}
\input{6-experiments}

\input{7-conclusion}

\clearpage

{\small
\bibliographystyle{plainnat}
\bibliography{ref}
}
\clearpage
\newpage
\appendix
\onecolumn
\input{appendix}

\end{document}

%% file: commands.tex
\usepackage{epsf}
\usepackage{graphics}
\usepackage{psfrag}

\usepackage{color}

\usepackage{mathtools}
\usepackage{amsfonts}
\usepackage{amssymb}

\newtheorem{proposition}{Proposition}

\long\def\comment#1{}

\newcommand{\bm}[1]{\boldsymbol{#1}}



%% file: 0-abstract.tex
\begin{abstract}
   The massive growth in the utilization of edge AI has made the applications of machine learning models ubiquitous in different domains. Despite the computation and communication efficiency of these systems, due to limited computation resources on edge devices, relying on more computationally rich systems on the cloud side is inevitable in most cases. Cloud inference systems can achieve the best performance while the computation and communication cost is dramatically increasing by the expansion of a number of edge devices relying on these systems. Hence, there is a trade-off between the computation, communication, and performance of these systems. In this paper, we propose a novel framework, dubbed as \Eccen that learns models with different levels of trade-offs between these conflicting objectives. This framework, based on an adaptation of knowledge from the edge model to the cloud one, reduces the computation and communication costs of the system during inference while achieving the best performance possible. The \eccen framework can be considered as a new form of compression method suited for edge-cloud inference systems to reduce both computation and communication costs. Empirical studies on classification and object detection tasks corroborate the efficacy of this framework.
   
\end{abstract}

%% file: 1-introduction.tex
\section{Introduction}\label{sec:intro}
\begin{figure}[t]
    \centering
    \includegraphics[width=0.5\columnwidth]{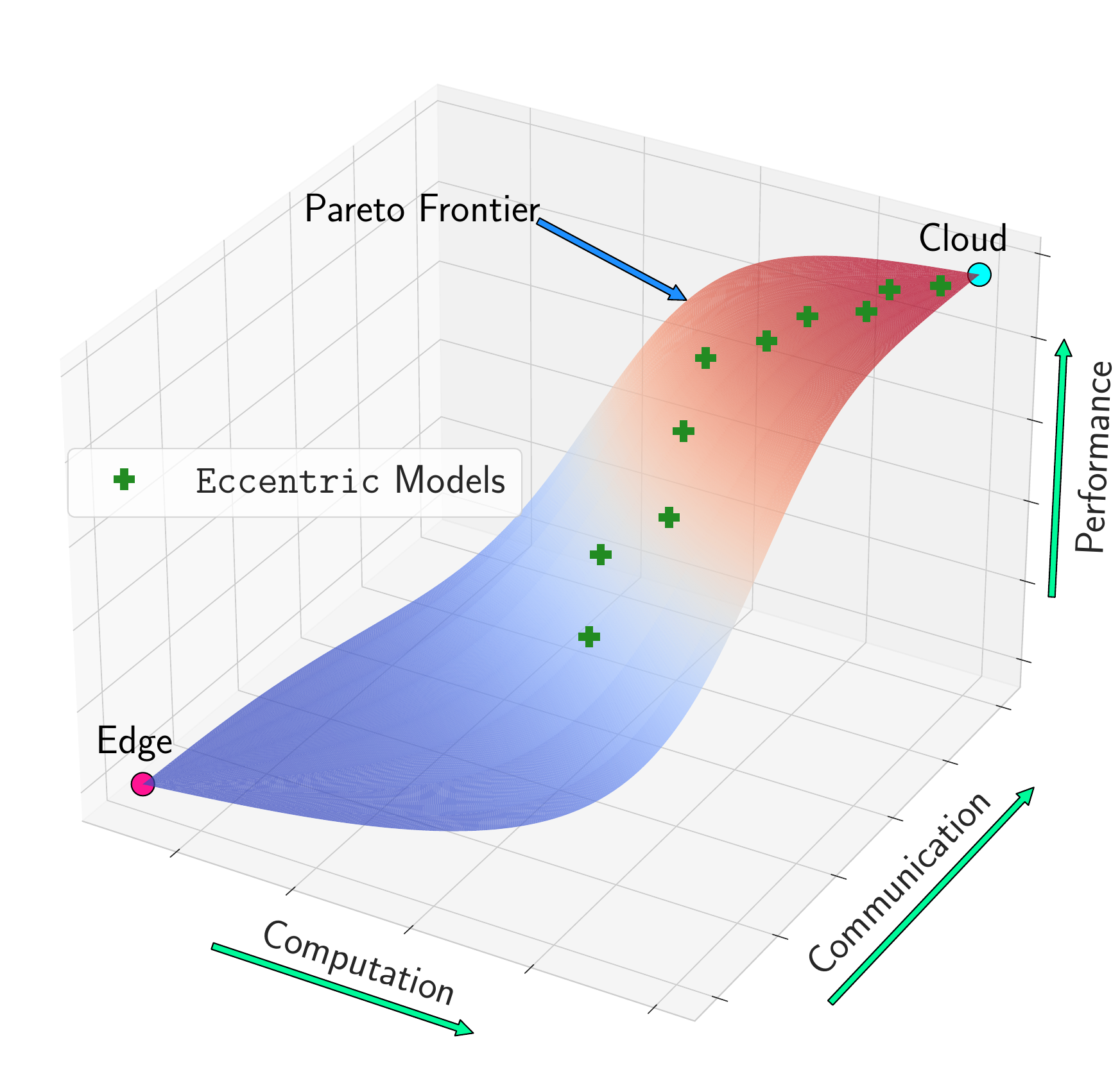}
    \caption{\Eccen framework seeks to fill the gap between the edge and cloud inference systems by learning Pareto optimal models with different levels of trade-off between computation, communication, and performance of the inference system from their Pareto frontier surface. Arrows show the direction of increase.}
    \label{fig:pareto}
    \vspace{-0.4cm}
\end{figure}

Nowadays, \textit{Edge AI} is playing a vital role in the advancement of machine learning and computer vision applications in numerous fields. State-of-the-art applications have been developed and deployed for these systems from object detection~\cite{bochkovskiy2020yolov4,yolov3,girshick2015fast,ren2015faster} and image classification~\cite{he2016deep,tan2019efficientnet,howard2017mobilenets} to natural language processing~\cite{mikolov2013efficient, vaswani2017attention} and generative models~\cite{goodfellow2014generative, vahdat2020nvae}. Notwithstanding these great achievements in different domains, the computational limitations of most of edge systems are the main hindrance of efficient and fast utilization of these models in those systems. The primary solution is to rely on a cloud system that has access to a higher computational resource in order to perform the inference more effectively. However, relying on a cloud system entails higher communication and computation costs. 

Hence, there is a conspicuous trade-off between \textit{Computation}, \textit{Communication}, and \textit{Performance} of these systems, where the edge and the cloud models are the two extremes in this trade-off. The edge model has the lowest computation and no communication cost, with possibly the lowest performance while the cloud model is the most performant model but at the cost of the highest computation and communication costs. Similar to other multi-criteria optimization problems~\cite{miettinen2012nonlinear}, when dealing with multiple conflicting objectives, we ought to seek points with optimal trade-offs, known as Pareto optimal points. The set of all Pareto optimal points is called Pareto frontier, as shown in Figure~\ref{fig:pareto} for the aforementioned trade-offs between these models. In this paper, we aim at introducing a new framework dubbed as \Eccen, to fill the gaps between edge-based and cloud-based inference systems by tracing some points on the Pareto frontier of this multi-criteria problem.


Many compression techniques such as knowledge distillation~\cite{hinton2015distilling,wang2019distilling,chennupati2021adaptivedistillationaggregatingknowledge} or quantization~\cite{rastegari2016xnor, polino2018model} have been proposed for the training of efficient and fast models; however, there is always a lower bound on the ability of these algorithms in reduction of computation without compromising the performance. Hence, they might not be sufficient to be able to run on low-powered edge devices, especially when the models are highly over-parameterized. In \eccen, we propose dynamic structures to work with the best model that can be run on the edge system, in conjunction with the cloud model to improve the performance of the edge model, while reducing computation and communication with respect to the cloud-based inference system. 
These models are based on adapting the knowledge at the edge level model to its counterpart on the cloud model, using techniques in knowledge distillation, but in a reverse direction from the student to the teacher. In addition, for our \eccen framework we propose using deep models for adaptation in knowledge distillation techniques, where we show it can improve distilling the knowledge from the teacher to the student as well. 

This dynamic structure of \eccen not only decides ``when'' to send the data to the cloud but also ``what'' should be sent. Despite the classical compression methods that are mainly focused on a fixed model structure to compress, \eccen models provide a dynamic structure that can adapt based on the input data. This dynamic structure efficiently reduces both computation and communication costs of edge-cloud systems, while trying to preserve the cloud model's performance. Hence, it can be considered as a new form of compression method, that optimizes communication costs in addition to the computation-performance trade-offs for an efficient inference system.
The contribution of this paper is summarized as follows:
\begin{itemize}
\item Investigating the computation, communication, and performance trade-offs of edge-cloud systems, and proposing \eccen framework to fill the gaps between the edge and cloud models on the Pareto frontier of this trade-off. Also, proposing new evaluation criteria for measuring these trade-offs.
\item Proposing the dynamic structure of \eccen framework as a new form of compression that considers communication costs in addition to computation and performance.
\item As a byproduct of the framework, we propose new adaptation modules for knowledge distillation to improve the distillation process.
\item Implementing the \eccen framework on two tasks of classification and object detection with empirical studies to corroborate the efficacy of \eccen framework.
\end{itemize}

%% file: 3-eccenteric.tex
\section{\Eccen Framework}\label{sec:ecc}

\begin{figure*}[t]
    \centering
    \includegraphics[width=0.98\textwidth]{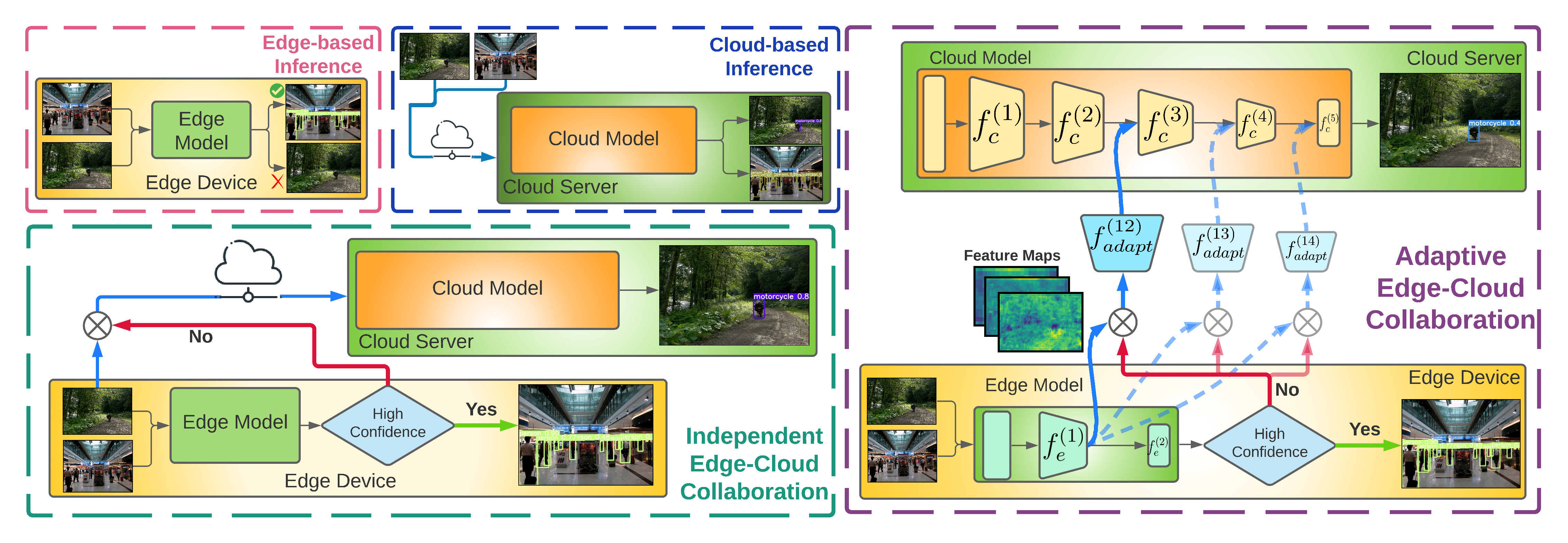}
    \caption{The proposed \Eccen frameworks for distributed inference using edge-cloud collaboration with knowledge adaptation.}
    \label{fig:Eccentric}
\end{figure*}
The ultimate goal of \eccen models is to reduce the communication and computation complexities of the cloud inference systems, while boosting the poor performance of edge inference systems. In another sense, we want to trace the models on the Pareto frontier of trade-offs between computation, communication, and performance of these models, in order to achieve the optimal compromises suitable for every scenario. This framework will give us more flexibility in choosing the right system based on resources available and targeted performance. We propose three different structures for \eccen framework and discuss how to train them.

\subsection{Problem Formulation}
In order to investigate the problem of distributed inference in deep neural networks, we first need to discuss the structure and learning process of these models. Consider that we have a deep neural network model $\bm{w} \in \mathbb{R}^{d}$, which consists of $M$ different layers $\bm{w} = \left\{\bm{w}_1,\ldots,\bm{w}_M\right\}$ (e.g. convolutional, fully-connected, residual, or any other layer represented by a parameter set of $\bm{w}_l, l\in [M]$).
Each layer takes its input as $\bm{x}_l, l\in [M]$, which is the output of the previous layer's forward processing using a function $\bm{y}_{l-1} = f_{l-1}\left(\bm{x}_{l-1};\bm{w}_{l-1}\right)$ (i.e. $\bm{x}_l = \bm{y}_{l-1}$).

Generally, in a supervised learning task such as classification or object detection, we want to optimize a prediction mapping on a training dataset $\mathcal{T}$, with $n$ training samples. The mapping transforms the feature space $\mathcal{X}$ to a label space $\mathcal{Y}$ (class label or object annotations), where each sample point is denoted by $(\bm{x}^{(i)},y^{(i)}) \in \mathcal{X}\times\mathcal{Y}$. This mapping can be represented with cascading of different layer functions of $f_l\left(\bm{x}_l^{(i)};\bm{w}_l\right)$, where $\bm{x}_l^{(i)}$ is the input of the $l$-th layer generated from the input sample $\bm{x}^{(i)}$. The set of all these functions is $\mathcal{F}\left(\bm{w}\right) \triangleq \left\{f_1\left(.;\bm{w}_1\right),\ldots,f_M\left(.;\bm{w}_M\right)\right\}$. Then, the goal is to minimize the empirical risk of training data on this model:
\begin{equation}\label{eq:mainloss}
    \mathcal{L}\left(\mathcal{F}\left(\mathcal{W}\right);\mathcal{T}\right) = \frac{1}{n}\sum_{i=1}^n\ell\left(\bm{w};\bm{x}_i,y_i\right),
\end{equation}
where $\ell(.;.,.)$ is the loss function for each sample data. 

The goal of either of the edge or the cloud models is minimize this empirical risk to achieve the best inference performance on test dataset, based on their models $\bm{w}_\text{edge}\in\mathbb{R}^{d^e}$ with $N_e$ layers, and $\bm{w}_\text{cloud}\in\mathbb{R}^{d^c}$ with $N_c$ number of layers, respectively. Due to the gap between the representational capabilities of these two models, their performance on the test dataset varies significantly. However, the limited computational power of the edge devices, which is the main bottleneck in such systems, would not allow filling this gap in the performance by increasing the edge model complexity. On the other hand, merely relying on the cloud model will cost substantially higher compared to edge-based systems due to both higher communication and computation resources required for these systems. Hence, the main goal of the \eccen framework is to minimize this loss by combining these models in a way to have the best performance possible with the lowest viable computation and communication with the cloud. 
\begin{equation}
    \mathcal{L}\left(\mathcal{F}_\ecc\left(.;\bm{w}_\ecc\right);\mathcal{T}\right) = \frac{1}{n}\sum_{i=1}^n\ell\left(\bm{x}_i,y_i; \bm{w}_\ecc\right),
\end{equation}
where $\mathcal{F}_\ecc \subset \mathcal{F}_\text{edge} \cup \mathcal{F}_\text{cloud}  \cup \mathcal{F}_\text{adapt}$, suggesting that the layers of the \eccen model are subset of the union of layers from the edge ($\mathcal{F}_\text{edge}$) and the cloud models ($\mathcal{F}_\text{cloud}$) as well as some \textit{adaptation} layers ($\mathcal{F}_\text{adapt}$) to connect the two models together. Note that, an \eccen model contains only a subset of those parameters and not all of them.

There are two primary characteristics of the edge AI systems that are the main motivation for introducing \eccen. First, in most edge AI systems, the generated sample data, such as video or images, usually, do not necessarily contain any targets of interest that the system is seeking to detect. These are mainly labeled as a normal class in the classification or background images in the object detection tasks. Hence, if the edge model can effectively detect these samples and filter them out before sending them to the cloud, we can significantly reduce the communication and computation of the cloud system. Second, even if all the data contain classes or objects to be detected, the edge model can handle parts of these detection tasks and pass the rest to the cloud side to reduce communication. Also, we can use the computed feature maps on the edge side and adapt them to the feature maps of the cloud model to bypass part of the cloud model and avoiding repeating the computation. We will discuss each of these structures and provide practical frameworks for different scenarios. 

\subsection{\Eccen for Distributed Inference}\label{sec:eccen}

The main idea behind \eccen is to distribute part of the inference on the edge side and complete the rest in the cloud systems. Thus, in some cases, the edge can perform the inference effectively, while in other cases it will use the cloud resources whenever necessary. In this case, the question is \textit{when} to send data to the cloud for a better inference using its resources. In addition to \textit{when}, we can ask \textit{what} should be sent to the cloud for further inference, considering that a part of inference has already done in the edge devices and the resulting output feature maps can be utilized for the cloud inference without sending the whole data itself. This strategy, not only is able to reduce the communication costs but also reduces the computation costs on the cloud side. Moreover, since the data samples are not directly sent to the cloud, it can protect the privacy of the data on the edge devices. We propose three different structures for inference using edge and cloud models involved in this framework. Using these variants of \eccen framework, we are able to train models with different levels of compromises in terms of communication, computation, and performance, that we can choose from based on the resources available in each system. 

\vspace{-2pt}\paragraph{\textbf{Independent \ecc}} In this structure, the edge model is used mainly as a filtration module to decide \textit{when} the input data should be sent to the cloud for further inference. This is based on the confidence that the edge device has in the inference output. In this structure, the input data will be sent as a whole to the cloud for inference, should the edge model decide to send it to the cloud. Two cases can be considered when the edge model can perform the inference by itself and do not pass the input data to the cloud. First, when the confidence of the edge model on the inference output for a sample is high. This could be class confidence in a classification task or the average of objects' confidence detected in an image for an object detection task. 

Another more important case happens when the edge device generates samples that do not contain any information we are seeking to detect. In this case, those samples do not contain any of the desired classes or objects we are looking for, and thereby, can be discarded by the edge to save computation and communication resources. This case is extremely common in most edge-based systems, in which most of the computation and communication resources are exhausted by such samples. From another point of view, this case can be considered as an instance of the first case described above. In this scenario, we can consider these samples as a separate class (i.e. normal class for classification task or background object in object detection task). Whenever the confidence of the model is high for this class or object it will conclude the inference, otherwise, send the image to the cloud for improved inference results. Thus, the independent \ecc model implements the following rule:

\begin{equation}\label{eq:ecci}
\mathcal{F}_{\texttt{\textbf{ECC}}_I}\left(\bm{x};\bm{w}_{\texttt{\textbf{ECC}}_I}\right) = \begin{cases}
\mathcal{F}_\text{edge}\left(\bm{x};\bm{w}_\text{edge}\right) &\mathsf{C}_\text{edge} \geq c_1\\
\mathcal{F}_\text{cloud}\left(\bm{x};\bm{w}_\text{cloud}\right) & \text{Otherwise},
\end{cases}
\end{equation}
where $\mathsf{C}_\text{edge}$ is the confidence of the edge model on the normal class or background object for their respective tasks and $c_1$ is the designed threshold. The schema of this framework is depicted in Figure~\ref{fig:Eccentric}.
An important note for this structure is that, since the edge model is mostly filtering the images, it is desired for this model to have as low \textit{false negative rate} as possible, not to miss any important sample. This requires to have a high recall rate in addition to accuracy. In Section~\ref{sec:training}, we use a multiobjective optimization approach to satisfy both objectives at the same time.

\vspace{-2pt}\paragraph{\textbf{Adaptive \ecc}} In this structure, the goal is to adapt the feature maps of the edge model to corresponding feature maps on the cloud model. In doing so, we use these adapted feature maps from the edge side as an input for the designated layers in the cloud model, and hence, bypass several layers in the cloud model, and lower computation costs overall. The output of the inference on the edge side is still used for filtration, but this time the feature maps are sent to the cloud instead of the input data. The adaptation process, using adaption modules occurs on the cloud side. In this structure, the training of the edge and cloud models, as well as adaptation modules, are coupled together. In Section~\ref{sec:training}, we will describe how to effectively learn to adapt the features to the cloud side to ensure that the adapted feature maps are as close as possible to the cloud's feature maps.

In adaptive \textbf{\texttt{ECC}}, we need to use modules to transfer the feature maps generated by the edge model to the corresponding cloud feature maps. These layers are denoted by $\mathcal{F}^{(mn)}_\text{adapt}$ and parameterized by $\bm{w}^{(mn)}_\text{adapt}$, where $m$ is the index of the feature map layer in the edge model and $n$ is the index of the feature map in the cloud model. These auxiliary layers adapt the output of the $m$-th layer in the edge ($\bm{y}^{(m)}_\text{edge}$) to the output of the $n$-th layer of the cloud model ($\bm{y}^{(n)}_\text{cloud}$):
\begin{align}\label{eq:adapt}
    \bm{y}^{(n)}_\text{adapt} = \mathcal{F}^{(mn)}_\text{adapt} \left(\bm{y}^{(m)}_\text{edge}; \bm{w}^{(mn)}_\text{adapt}\right).
\end{align}
The objective is to minimize the distance between feature maps of $\bm{y}^{(n)}_\text{adapt}$ and $\bm{y}^{(n)}_\text{cloud}$, where we use knowledge distillation approaches during the training to achieve this goal. During the inference, similar to independent \ecc, we use a threshold $c_1$ to filter samples, but instead of the whole image, we will send the feature maps to the cloud:
\begin{equation}\label{eq:ecca}
\mathcal{F}_{\texttt{\textbf{ECC}}_\text{A}}\!\!\left(\bm{x};\bm{w}_{\texttt{\textbf{ECC}}_\text{A}}\right) \!\!= \!\!
\begin{cases}
\mathcal{F}_\text{edge}\left(\bm{x};\bm{w}_\text{edge}\right) &\mathsf{C}_\text{edge} \geq c_1\\
\mathcal{F}^{(l>n)}_\text{cloud}\left(\bm{y}^{(n)}_\text{adapt};\bm{w}^{(l>n)}_\text{cloud}\right) & \text{Otherwise},
\end{cases}
\end{equation}
where $\bm{y}^{(n)}_\text{adapt}$ is calculated from the input data and the resulting edge feature map using (\ref{eq:adapt}), and $\mathcal{F}^{(l>n)}_\text{cloud} = \left\{f^{(n)}_\text{cloud}\left(.;\bm{w}^{(n)}_\text{cloud}\right),\ldots,f^{(N_c)}_\text{cloud}\left(.;\bm{w}^{(N_c)}_\text{cloud}\right)\right\}$ and $\bm{w}^{(l>n)}_\text{cloud} = \left\{\bm{w}^{(n)}_\text{cloud},\ldots, \bm{w}^{(N_c)}_\text{cloud}\right\}$ are layer functions and corresponding parameters after the $n$-th layer in the cloud model. 
A schema of this framework is shown in Figure~\ref{fig:Eccentric}.

\vspace{-2pt}\paragraph{\textbf{Dynamic \ecc}} The independent \texttt{\textbf{ECC}}, as we will show in Section~\ref{sec:exp}, can perform as good as the cloud model counterpart, however, the computation cost is still a burden and might delay the inference time since it needs the data to pass through both the edge and the cloud models for some inputs. On the other side, the Adaptive \ecc can efficiently reduce the computation costs by sacrificing some performance measures compared to the cloud model. Hence, a better idea is to combine them together and benefit from both worlds. To do so, we need to find a mechanism to decide when to use each of these models dynamically on different input data. An approach would be to use the confidence level of the inference result on the edge side to decide between these two structures for each input data. An ablation study in Section~\ref{sec:exp}, confirms that the adaptive \eccs performance degrades as the edge model's confidence decreases on an input data. We will show that using Dynamic \ecc we can trace different points on the Pareto frontier of the trade-offs between computation, communication, and performance of edge-cloud models. Hence, by finding the best threshold for this transition, the structure of this model is:
\begin{equation}\label{eq:ecca}
\mathcal{F}_{\texttt{\textbf{ECC}}_\text{D}}\!\!\left(\!\bm{x};\bm{w}_{\texttt{\textbf{ECC}}_\text{D}}\!\right) \!\!=\!\! \begin{cases}
\mathcal{F}_\text{edge}\left(\bm{x};\bm{w}_\text{edge}\right) &\!\!\!\!\!\!\mathsf{C}_\text{edge} \geq c_1\\
\mathcal{F}^{(l>n)}_\text{cloud}\!\!\left(\bm{y}^{(n)}_\text{adapt};\bm{w}^{(l>n)}_\text{cloud}\right) & \!\!\!\!\!\!c_2 \leq \mathsf{C}_\text{edge} < c_1\\
\mathcal{F}_\text{cloud}\left(\bm{x};\bm{w}_\text{cloud}\right)  & \!\!\!\!\!\!\mathsf{C}_\text{edge} < c_2
\end{cases}
\end{equation}

%% file: 4-training.tex
\section{Training Strategies}\label{sec:training}
To train each of these structures, we use different techniques in order to boost the performance close to the cloud model's while reducing the communication and computation cost overall. These methods are mostly used for training adaptive \ecc models. We will first describe knowledge distillation and knowledge adaptation and the training algorithm for an adaptive \ecc model.

\vspace{-2pt}\paragraph{Knowledge Distillation} Boosting the performance of a student model using a teacher model has been widely explored in different studies~\cite{hinton2015distilling,wang2019distilling,chen2017learning}, with various approaches. In all these approaches, the goal is that the student can perform independently from the teacher after the training. In addition, the boost in the performance of the student model using this approach  is largely confined to the complexity of the student model, and hence, very limited. However, in the proposed \eccen framework, we aim at using the cloud model in conjunction with the edge model to boost the performance of the edge model close to its cloud counterpart. To do so, we will employ the cloud model to distill the knowledge to the edge model, and then, will use the cloud model or part of it after the training in one of the proposed \eccen structures.

There are two main approaches for knowledge distillation used in the literature: (1) Distilling knowledge through the confidence scores by adjusting the temperature in the Softmax function~\cite{hinton2015distilling}, and (2) Distilling the knowledge using hint layers and feature imitation for different layers of a neural network~\cite{romero2014fitnets,wang2019distilling}. In \eccen framework, we are focusing on the latter form since it can be used for knowledge adaptation part as well. However, the former approach still can be used to boost the performance of the edge model. For the hint layers, most of the current solutions are using a simple adaptation module such as one layer fully-connected network~\cite{romero2014fitnets} or a layer of convolution network~\cite{wang2019distilling}, to mainly focus on the student model's parameters rather than the adaptation module. However, in this work we are proposing the use of deep networks such as residual layers~\cite{he2016deep} or CSP bottleneck layers~\cite{wang2020cspnet}, similar to networks used in domain adaptation~\cite{long2016unsupervised} and variational autoencoders~\cite{vahdat2020nvae}, for two main reasons. First, we will show that using these deep networks we can boost the performance of the student model better than those simple networks. Second, we will use the adaptation modules for our knowledge adaptation part, and using a deep model can achieve a better performance. In Section~\ref{sec:exp} and Appendix~\ref{app:add_exp}, we will further elaborate on the structure of adaptation module we used for each task. To distill the knowledge from the cloud model, we use the binary cross-entropy loss between the adapted feature map of the $m$-th layer of the edge model using (\ref{eq:adapt}) and the cloud's feature map on the $n$-th layer:
\begin{align}
    \mathcal{L}_\text{KD} \left(\mathcal{F}_\ecc;\mathcal{T}\right) =
    \mathsf{BCE}\left(\sigma\left(\bm{y}^{(n)}_\text{cloud}\right),\sigma\left(\bm{y}^{(n)}_\text{adapt}\right) \right),
\end{align}
where the $\sigma(x) = 1/(1+e^{-x})$ is the Sigmoid function to normalize the values of each pixel in the feature maps of the cloud and adaptation models. This loss will update adaptation module parameters $\bm{w}^{(mn)}_\text{adapt}$, as well as the edge model parameters on and before the $m$-th layer $\bm{w}^{(l\leq m)}_\text{edge}$. In conjunction with the main learning objective defined in (\ref{eq:mainloss}) for the edge model, we can optimize the edge and adaptation model parameters. In Appendix~\ref{app:add_exp}, we will show the effectiveness of these deep layers for knowledge distillation over single layers used by other approaches.

\vspace{-2pt}\paragraph{Knowledge Adaptation}
After using knowledge distillation, as described above to boost the performance of the edge model, we keep the learned adaptation modules to be used in our adaptive \ecc framework. To further optimize these parameters, we run a fine-tuning stage using only the adaptation parameters $\bm{w}^{(mn)}_\text{adapt}$ and the rest of the cloud model parameters used in the adaptive \ecc model, $\bm{w}^{(l>n)}_\text{cloud}$. Due to lack of space, the details of the algorithm for \ecca model, as well as more experimental results are deferred to the Appendix.

\vspace{-2pt}\paragraph{Recall rate Boosting} As it was mentioned, in most cases the edge model is used as a filtration module for an \ecc model, to only pass samples with desired classes or objects for improved inference to the cloud. However, this requires a high recall rate for the edge model, not to miss any important sample, in addition to the main objective of the training for the edge model. To do so, we need to add another objective, which is the loss only on the positive samples\footnote{Positive samples are samples with classes other than normal in the classification and samples with at least one desired object in the object detection task.}. This loss will ensure to have a higher recall rate, but it might be at odds with the main training objective to some extent (decreasing precision rate), hence, we need to use multiobjective optimization approaches~\cite{miettinen2012nonlinear,cortes2020agnostic}, to ensure that both objectives are minimized simultaneously. To do so, we can reweight the objectives to converge to a point from the Pareto frontier of the problem. Instead of manually assigning the weights, we use the following proposition that ensures to find the best weights at every step to have a decreasing gradient direction for all objectives. 
\begin{proposition}[Pareto Descent Direction]
Consider a multiple objective problem with $p$ objectives of $\bm{\mathrm{h}}\left(\bm{w}\right) = \left[\mathrm{h}_1\left(\bm{w}\right), \mathrm{h}_2\left(\bm{w}\right), \ldots, \mathrm{h}_p\left(\bm{w}\right)\right]$, that ought to be minimized. Using the solution of the following quadratic optimization as the weights for each objective:
\begin{equation}\label{eq:quad}
    \bm{\alpha}^*(\bm{w}) \in \arg\underset{\bm{\alpha} \in \Delta_p}{\min} \left\|\sum_{i=1}^p \alpha_i\bm{\mathrm{g}}_i\left(\bm{w}\right)\right\|_2^2,
\end{equation}
where $\bm{\mathrm{g}}_i\left(\bm{w}\right) = \nabla_{\bm{w}}\mathrm{h}_i\left(w\right), i\in [p]$ and $\Delta_p$ is a $p$-dimensional simplex. We can show that the gradient of the weighted sum of objectives, using the solution of (\ref{eq:quad}) as the weights, is either zero or a descent direction for all the objectives. Meaning, we have:
\begin{equation}\label{eq:paretcond}
    - \left\langle \sum_{i\in [p]} \alpha^*_i \bm{\mathrm{g}}_i\left(\bm{w}\right), \bm{\mathrm{g}}_j(\bm{w}) \right\rangle\leq 0, \;\; \forall j\in\{1,\ldots,p\}\;.
\end{equation}
\label{prop:pareto}
\end{proposition}
This proposition will ensure that at every step of the gradient descent we will not increment any of the objectives until reaching a point from the Pareto frontier. The proof of this proposition is deferred to Appendix~\ref{app:pareto}.

%% file: 5-eval.tex
\section{Evaluation Criteria}\label{sec:eval}
To fairly evaluate models trained by different structures of the \eccen framework, we need to consider all three objectives of computation, communication, and performance. To do so, we define the following metrics to have a fair comparison between these models with respect to corresponding metrics on the edge and cloud models. Hence, we define three main scores for each objective as follows:

\vspace{-2pt}\paragraph{Communication Score ($\mathsf{S}_\text{comm}$)} This score defines how much communication is needed for any defined \eccen model compared to a fully cloud-based system. It consists of two parts: First is the throughput of the data communicated to the cloud as the ratio between number of communicated samples ($N_p$) over total number of samples ($N$): $\tau = {N_p}/{N} \in \left[0,1\right]$.
Second is the size of communication, indicated by the ratio between the size of the data communicated for each sample to the cloud ($|\mathcal{D}_p|$), over the original size of the input sample ($|\mathcal{D}|$): $\psi = {|\mathcal{D}_p|}/{|\mathcal{D}|} \in \mathbb{R}_+$. Then, the communication score will be defined as:
$
    \mathsf{S}_\text{comm} \triangleq \tau \cdot \psi.
$
The score is zero for the edge model, and it is one for the cloud model that has the maximum communication.

\vspace{-2pt}\paragraph{Computation Score ($\mathsf{S}_\text{comp}$)} To measure the computation complexity between different models, we will use their floating point operation counts as ($\mathsf{FLOPS}$).
\begin{align}
    \mathsf{S}_\text{comp} \triangleq \frac{\mathsf{FLOPS}\left(\bm{w}_\ecc\right) - \mathsf{FLOPS}\left(\bm{w}_\text{edge}\right)}{\mathsf{FLOPS}\left(\bm{w}_\text{cloud}\right) - \mathsf{FLOPS}\left(\bm{w}_\text{edge}\right)}.
\end{align}
This definition will ensure that in this scale the computation score of the edge model is zero while that of the cloud model is one. Note that, in calculation the $\mathsf{FLOPS}$ for the \ecc model we are dealing with two parts, one is the edge section and the other is the part resides in the cloud. All datapoints are passing through the edge model ($\bm{w}_\text{edge}$) while only part of them (with ratio of $\tau$) is passed through the cloud side model ($\bm{w}^c_\ecc$). Hence, for calculating the computation complexity of the \ecc model we use:
$
\mathsf{FLOPS}\left(\bm{w}_\ecc\right) = \mathsf{FLOPS}\left(\bm{w}_\text{edge}\right) + \tau\mathsf{FLOPS}\left(\bm{w}^c_\ecc\right).
$
\vspace{-2pt}\paragraph{Performance Score ($\mathsf{S}_p$)} There is a gap between the performance of the edge and cloud models, and this score wants to show how much of this gap is filled with the trained \eccen model. To do so, for any performance metric $\pi$ (e.g. accuracy, recall rate, or mAP), we can define the performance score of an \eccen model as:
\begin{align}
    \mathsf{S}_p \triangleq \frac{\pi_\texttt{ECC} - \pi_\text{edge}}{\pi_\text{cloud} - \pi_\text{edge}},
\end{align}
where, $\pi_\texttt{ECC}$, $\pi_\text{edge}$, and $\pi_\text{cloud}$ are the performance measures of the \ecc, edge and cloud models respectively. Based on this definition, the performance score of the edge model is zero and that of the cloud model is one. It can be even greater than one, which means the trained \ecc model is performing better in that specific performance measure than the cloud one (as shown for some cases in Section~\ref{sec:exp}).

%% file: 6-experiments.tex
\section{Experiments}\label{sec:exp}
In this section, we evaluate the performance of the proposed \ecc models on two different tasks of classification and object detection and show their efficacy in practice. More details on the experiment, as well as more results, are deferred to the Appendix~\ref{app:add_exp} due to the space limit.
\vspace{-2pt}\paragraph{Experimental Setup} For the classification task, we use the CIFAR10 dataset~\cite{krizhevsky2009learning}. We modify the classes to be similar to the edge-cloud setting described in Section~\ref{sec:eccen}, where the objective is only to classify certain classes and all the rest are treated as the normal class. For this purpose, we consider ``animal'' classes (i.e. bird, cat, deer, dog, frog, and horse), and consider all non-animal classes (i.e. airplane, automobile, ship, truck) as one class of normal cases. Hence, the classification is among $7$ classes instead of $10$. For the cloud model we use ResNet18~\cite{he2016deep}, where it contains $4$ residual layers, each of which consists of $2$ residual blocks. For the edge model, we use a model we call TinyResNet, which contains one residual layer with $2$ residual blocks. Details are explained in the Appendix~\ref{app:add_exp}.

For the object detection, we use COCO dataset~\cite{lin2014microsoft}, which contains around $120$K of training images and $5$K of validation images with $80$ objects. Similar to classification, we only consider top $4$ objects in this dataset, which includes ``person'', ``bicycle'', ``car'', and ``motorcycle''. About $60\%$ of images in the COCO dataset contain at least one of these four objects. We consider the rest of the images as the background images. For the object detection, we use a variant of a \textsf{YOLO} SSD model~\cite{redmon2016you,redmon2016yolo9000,yolov3,bochkovskiy2020yolov4}, named \textsf{YOLOv5}\footnote{https://github.com/ultralytics/yolov5}. For the cloud model, we use the \textsf{YOLOv5l}, which is the large model provided by the repository. For an input image with a size of $(512,512,3)$ this model requires about $37.12~ \mathsf{GFLOPS}$ to process, and it has $91.6$ MB of parameter size. For the edge model, we train our model base on the \textsf{YOLOv5} structure named as \textsf{YOLOv5xs}. For the same input size, it requires about $0.35~ \mathsf{GFLOPS}$ to process, with $1.1$ MB of parameter size. All the computation scores in this section are computed based on the input size of $(32,32,3)$ for the classification task and $(512,512,3)$ for the object detection task.
\subsection{Main Results}
\vspace{-2pt}\paragraph{Classification Task} In the classification task, the goal is to classify images with an animal in them or discard those that do not contain any animal in it. For $\textbf{\texttt{ECC}}_\text{A}$, we use residual blocks as our adaptation modules, with varying number of residual layers for different models. We show these models with $\mathrm{r}_i, i\in\{1,2,3\}$, where the index shows the number of residual layers. For these models, we adapt the feature map of the first and only residual layer in the edge model to either of feature maps from $2$-nd, $3$-rd, or $4$-th residual layers in the cloud model. For more details on the adaptation structure refer to Appendix~\ref{app:add_exp}.

\begin{table}[t]
    \centering
    \resizebox{0.7\columnwidth}{!}{
    \begin{tabular}{llll}
    \toprule
                    &  \multicolumn{3}{c}{Objectives}
        \\ \cmidrule(r){2-4}
         Model        & Accuracy ($\mathsf{S}_p$)      & Compute ($\mathsf{S}_\text{comp}$) &  $\mathsf{S}_\text{comm}$
        \\
        \midrule
         \makecell[l]{Cloud \\ (ResNet18) } & $91.83\%$ \Chartgt{1.00}  & $38.50$ \Chartrt{1.00} & \Chartr{1.00}
        \\
        \midrule
          \makecell[l]{Edge \\ (ResNet5) }&  $77.32\%$ \Chartgt{0.00}& $3.47~~$ \Chartrt{0.00}& \Chartr{0.00}
        \\
        \midrule
        $\textbf{\texttt{ECC}}_\text{I}$& $91.01\%$ \Chartgt{0.9435}& $26.88$ \Chartrt{0.6682}& \Chartr{0.61}\\
        \midrule
        \makecell[l]{$\textbf{\texttt{ECC}}_\text{A}$ \\ (Best ACC)} & $90.92\%$ \Chartgt{0.9373}& $23.52$ \Chartrt{0.5713}& \Chartr{0.80}\\
        \makecell[l]{$\textbf{\texttt{ECC}}_\text{A}$ \\ (Best Compute)} & $84.80\%$ \Chartgt{0.5155}& $5.25~~$ \Chartrt{0.084}& \Chartr{0.80}\\
        \midrule
        \makecell[l]{$\textbf{\texttt{ECC}}_\text{D}$\\ (Best ACC)} & $91.33\%$ \Chartgt{0.9655}& $26.81$ \Chartrt{0.6665}& \Chartr{0.81}\\
        \makecell[l]{$\textbf{\texttt{ECC}}_\text{D}$\\ (Best Compute)} & $85.41\%$ \Chartgt{0.5575}& $6.57~~$ \Chartrt{0.0886}& \Chartr{0.80}\\
        \bottomrule
    \end{tabular}
    }
\caption{The trade-offs between performance (accuracy), computation and communication of different models. Bars show the corresponding scores. Green bars: the higher value is better. Red bars: the lower is better. Computations are in $\mathsf{MFLOPS}$.}
\label{table:classification}
\end{table}
\begin{figure}[t]
    \centering
    \includegraphics[width=0.95\columnwidth]{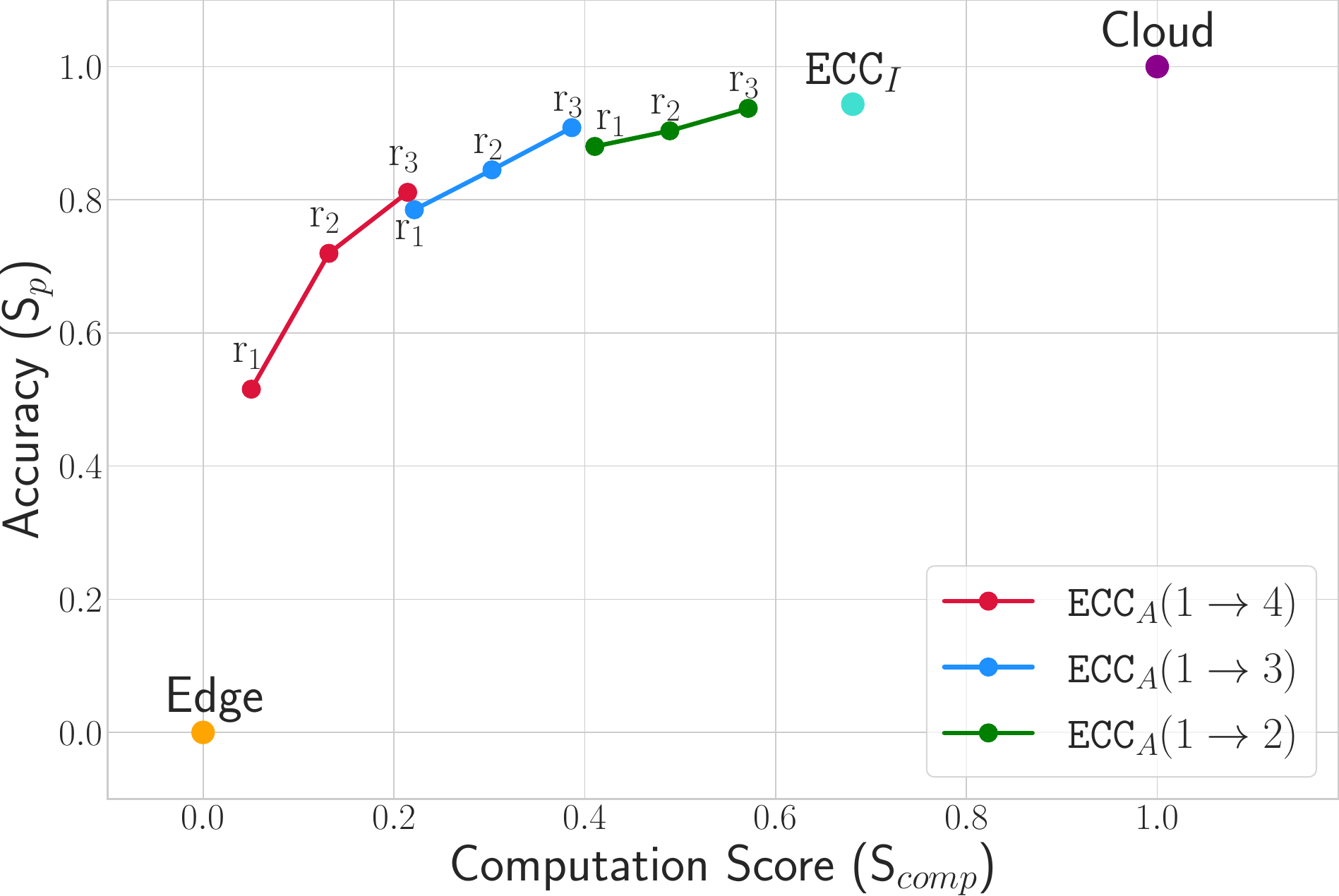}
    \caption{Trade-off between computation and performance of different \ecc models. The \ecca models are adapting the first residual layer of the edge model to either of the last three layers of the cloud model (2,3,4), by different adaptation modules with different number of layers $\{\mathrm{r}_1,\mathrm{r}_2,\mathrm{r}_3\}$}.\vspace{-0.05in}
    \label{fig:class_comp_perf}
\end{figure}

Table~\ref{table:classification} summarizes the results of these models comparing with their edge and cloud counterparts. As it can be inferred from Table~\ref{table:classification}, \ecci can achieve almost the same accuracy as the cloud model with $34\%\!\downarrow$ reduction in the computation of the system, and $39\%$ reduction in the overall communication. \ecca can achieve almost the same accuracy with reducing $43\%\!\downarrow$ of computation and $20\%\!\downarrow$ of communication. The communication is higher than \ecci since the feature maps have a higher size than the input data. This is not the case for all models, especially in deeper models. For instance, in the object detection task, the size of feature maps sent to the cloud is less than $1/3$ of the size of the input. The \eccd models with tuning the threshold can achieve the best accuracy among all \ecc models with $34\%\!\downarrow$ reduction in computation and $19\%\!\downarrow$ reduction in communication size. Figure~\ref{fig:class_comp_perf} shows the trade-off between the performance score and computation score of these models with respect to cloud and edge inference systems. The \ecca models are adapting from the residual layer of the edge model to either of the last three residual layers in the cloud model with a different number of residual layers as the adaptation module.

\begin{table*}[t]
    \centering
    \resizebox{0.75\textwidth}{!}{
    \begin{tabular}{llllll}
    \toprule
                    &  \multicolumn{5}{c}{Objectives}
        \\ \cmidrule(r){2-6}
         Model        & $\mathsf{mAP}@0.5$ ($\mathsf{S}_p$) & $\mathsf{mAP}@0.5\!\!:\!\!0.95$ ($\mathsf{S}_p$) & $\mathsf{F1}$ ($\mathsf{S}_p$)     & Compute ($\mathsf{S}_\text{comp}$) &  $\mathsf{S}_\text{comm}$
        \\
        \midrule
         \makecell[l]{Cloud \\ (\textsf{YOLOv5l}) } & $0.777$ \Chartgt{1.00}  & $0.552$ \Chartgt{1.00} & $0.541~$\Chartgt{1.00} & $37.12$ \Chartrt{1.00}& \Chartr{1.00}
        \\
        \midrule
          \makecell[l]{Edge \\ (\textsf{YOLOv5xs}) }& $0.454$ \Chartgt{0.00}  & $0.224$ \Chartgt{0.00} & $0.313~$\Chartgt{0.00} & $0.35~~$ \Chartrt{0.00}& \Chartr{0.00}
        \\
        \midrule
        \ecci& $0.776$ \Chartgt{0.997}  & $0.554$ \Chartgt{1.006} & $0.555~$\Chartgt{1.076} & $23.99$ \Chartrt{0.643}& \Chartr{0.64} \\
        \midrule
        \makecell[l]{$\textbf{\texttt{ECC}}_\text{A}(\mathsf{B}\rightarrow\mathsf{B},b_4)$ \\ (Best mAP)} & $0.661$ \Chartgt{0.640}  & $0.433$ \Chartgt{0.637} & $0.443~$\Chartgt{0.57} & $14.63$ \Chartrt{0.382}& \Chartr{0.19}\\
        \makecell[l]{$\textbf{\texttt{ECC}}_\text{A}(\mathsf{H}\rightarrow\mathsf{H},b_4)$ \\ (Best Compute)} &$0.522$ \Chartgt{0.2046}  & $0.291$ \Chartgt{0.1792} & $0.362~$\Chartgt{0.1815} & $3.86~~$ \Chartrt{0.090}& \Chartr{0.18}\\
        \midrule
        \makecell[l]{$\textbf{\texttt{ECC}}_\text{D}(c_2=0.45)$ \\ (Best mAP)}  &  $0.771$ \Chartgt{0.9825}  & $0.548$ \Chartgt{0.9867} & $0.553~$\Chartgt{1.053} & $21.31$ \Chartrt{0.5702}& \Chartr{0.54}\\
        \makecell[l]{$\textbf{\texttt{ECC}}_\text{D}(c_2=0.2)$ \\ (Best Compute)} &$0.572$ \Chartgt{0.3655}  & $0.336$ \Chartgt{0.3407} & $0.396~$\Chartgt{0.3636} & $7.65~~$ \Chartrt{0.1987}& \Chartr{0.27}\\
        \bottomrule
    \end{tabular}
    }
\caption{The trade-offs between performance, computation and communication of different models for object detection. Bars show the corresponding scores. In the green bars the higher value is better while in the red bars the lower is better. Computations are in $\mathsf{GFLOPS}$.}\vspace{-0.05in}
\label{table:object}
\vspace{-0.25cm}
\end{table*}

\begin{figure}[t]
    \centering
    \includegraphics[width=0.9\columnwidth]{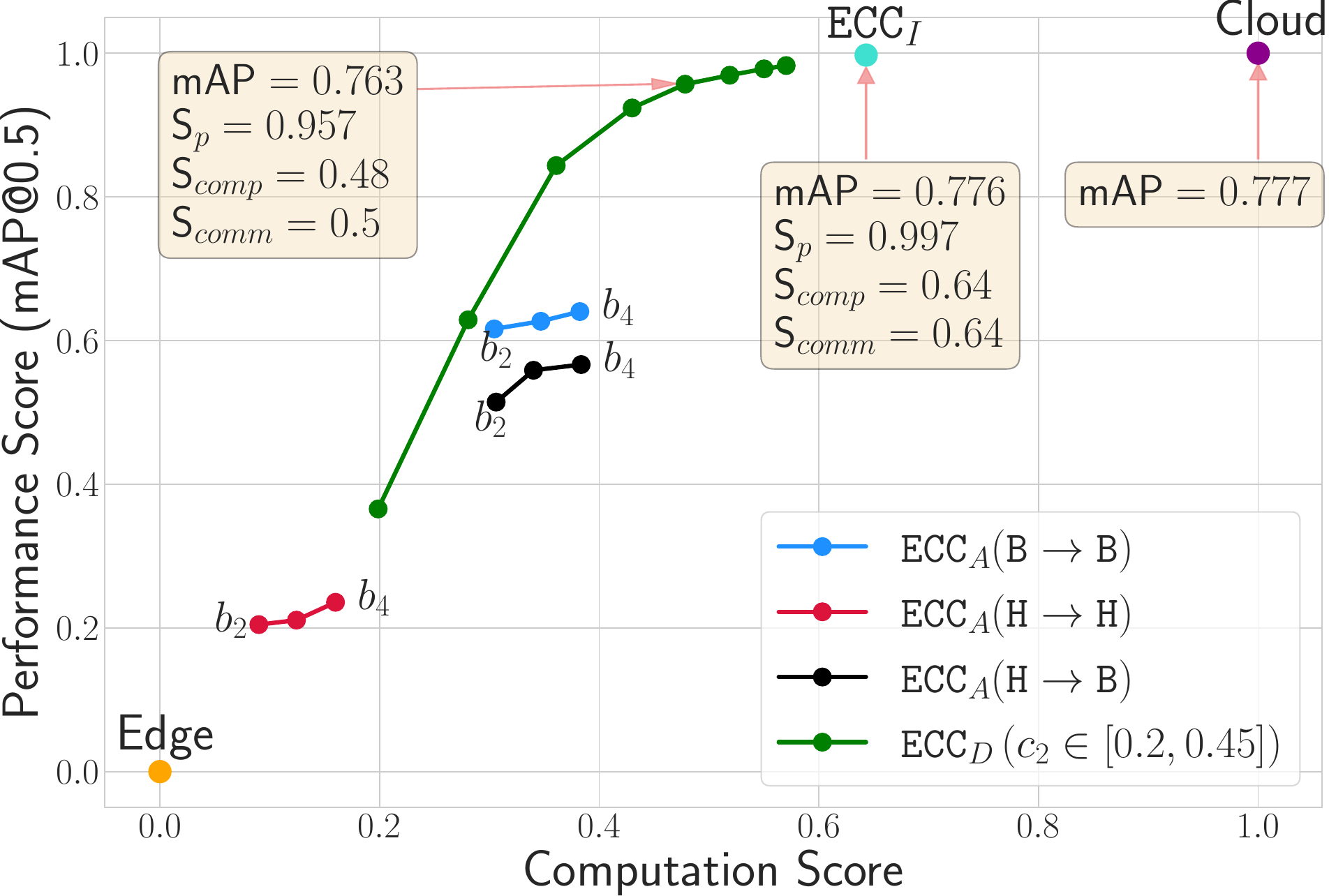}
    \caption{Trade-off between computation and performance of different \ecc models for object detection. The \eccd models achieve the same level of performance as the cloud with greater than $40\%$ reduction in computation and communication.}.
    \label{fig:object_comp_perf}\vspace{-0.25cm}
\end{figure}

\vspace{-2pt}\paragraph{Object Detection Task} For the object detection task we use \textsf{YOLOv5} models, where we can adapt the feature maps from their ``backbone'' ($\mathsf{B}$) or ``head'' ($\mathsf{H}$) modules for \ecca models. We consider three different forms of adaptation, that is, from backbone of the edge model to the backbone of the cloud model ($\mathsf{B} \rightarrow \mathsf{B}$), from the head to the backbone ($\mathsf{H} \rightarrow \mathsf{B}$), and from the head to the head ($\mathsf{H} \rightarrow \mathsf{H}$). For adaptation module, we consider using bottleneck layer~\cite{wang2020cspnet}, with three different sizes of $2,3,$ and $4$ layers, which we call them $b_2,b_3,$ and $b_4$. Also, for the \eccd models, we mainly use the trained models for $\textbf{\texttt{ECC}}_\text{D}(\mathsf{H} \rightarrow \mathsf{H}, b_2)$ as the base model and change the $c_2$ parameter defined in (\ref{eq:ecca}) in the range of $\left[0.2,0.45\right]$. Then we will only keep non-dominated models of \eccd that can generate the Pareto frontier as it is depicted in Figure~\ref{fig:object_comp_perf}. 

In Table~\ref{table:object}, a summary of training \ecc models for this object detection task is provided. As it can be inferred, the \ecci model can achieve the same performance of the cloud model (with even slightly higher $\mathsf{mAP}@0.5\!\!:\!\!0.95$ and $\mathsf{F1}$ measures) with $36\%\!\downarrow$ reduction in the computation and $36\%\!\downarrow$ reduction in the communication, which means that it can even dominate the cloud model in this trade-off and replace it. In \ecca models, the drop in performance increases with $35\%$ decrease (still $65\%$ higher than the edge model in this scale) in the performance for $\textbf{\texttt{ECC}}_\text{A}(\mathsf{B}\rightarrow\mathsf{B},b_4)$, while the computation and communication scores are significantly dropping by $62\%\!\downarrow$ and $81\%\!\downarrow$ respectively,  with respect to the cloud model. \eccd models can give us ranges of different models with various levels of trade-offs between three objectives. Its best model in terms of performance can achieve the same level of performance as the cloud with an even higher drop in computation and communication than \ecci ($43\%\!\downarrow$ and $46\%\!\downarrow$ decrease respectively). Figure~\ref{fig:object_comp_perf} shows the trade-off between performance and computation in these models. More comparisons can be found in the Appendix. It can be inferred that models generated by \ecca for $\mathsf{B}\rightarrow\mathsf{B}$ and $\mathsf{H}\rightarrow\mathsf{B}$ are dominated by the \eccd models, and hereby, cannot be on the Pareto frontier curve of this trade-off. It also can be seen that with an \ecca model with only $52\%$ of computation and $50\%$ of communication of the cloud inference system, we can achieve more than $95\%$ of the cloud model's performance ($\mathsf{mAP}@0.5=0.763$).

%% file: 7-conclusion.tex
\section{Conclusion and Future Works}\label{sec:conc}
In this paper, we propose \Eccen framework with three different structures to learn models with optimal trade-offs between computation, communication, and performance, in order to fill the gaps between edge- and cloud-based inference systems. We use knowledge adaptation to send the adapted feature maps instead of input data and greatly reduce the computation costs on the cloud side. Our empirical studies indicate that by learning these \ecc models with optimal trade-offs we can achieve the same level of performance as the cloud model while notably reducing the communication and computation costs. In addition, we show that by using deep structures for the adaptation modules we can improve the knowledge distillation process over classical approaches.

This framework can be generalized to other tasks beyond classification and object detection, which will be left as a future direction. Moreover, feature adaptation mechanism using the proposed adaptation modules can be improved with generative models in future works.

%% file: appendix.tex

\input{2-related}

\section{Additional Experimental Results}\label{app:add_exp}
In this section, we elaborate on more details of the empirical studies for our proposed \Eccen framework. We first provide more detailed results in addition to what is provided in the main body. Then, we go into the details of the model structures used in Section~\ref{sec:exp}.

\subsection{Additional Results}
\paragraph{Knowledge Distillation} In the main body, we mention that knowledge distillation using hint layers introduced by~\citet{romero2014fitnets} can be improved by employing deep architectures as the adaptation modules. In this part, we want to show empirically with an ablation study that using deep architectures such as residual blocks or bottleneck CSP modules can further improve the process of distillation. First, to showcase this idea, we use the classification task introduced in Section~\ref{sec:exp}. In this ablation study, we use the TinyResNet as the edge or student model and use the ResNet18 as the teacher model. Here, we use the edge model as a binary classifier between animal and non-animal classes of the CIFAR10 dataset. Then we use the adaptation module to distill the knowledge from the last residual layer of the teacher model (with the size of $\left(128,4,4\right)$ for the input size of $\left(32,32,3\right)$) to the output of the only residual layer of the edge model (with the size of $\left(16,16,16\right)$). As for the adaptation, we first use a simple Conv layer that has the input channel of $16$ and output channel of $128$, with the stride of $4$. Then, we utilize residual layers with different sizes of $1$,$2$, and $3$ layers to investigate the effect of deep residual layers and the number of residual layers in the distillation process. We run the training using knowledge distillation for $80$ epochs, where it converges. In Table~\ref{table:kd-class} the results of this ablation study are summarized, where it can be inferred that using deep architectures such as residual layers is more effective than simple adaptation modules. However, the depth of this adaptation module is important as well. As it can be seen, the best results are achieved with $2$ layers of residual modules. It is worth mentioning that, this ablation only considers the effects of distillation from one layer to another using hint layers and not any other forms of distillation. Using other forms, as well as using additional hint layers for the distillation can definitely further improve the performance over the baseline.

\begin{table}[t]
    \centering
    \begin{tabular}{llll}
    \toprule
                    &  \multicolumn{3}{c}{Objectives}
        \\ \cmidrule(r){2-4}
         Model      & Test Accuracy & Recall Rate  & Test Loss
        \\
        \midrule
         TinyResNet  & \makecell[c]{$93.421\%$ \\ $\pm 0.102\%$}  &\makecell[c]{$94.23\%$ \\ $\pm 0.526\%$} & \makecell[c]{$0.17$ \\ $\pm 0.002$}
        \\
        \midrule
          \makecell[l]{TinyResNet-KD \\ (with Conv) }&\makecell[c]{$93.556\%$ \\ $\pm 0.057\%$} & \makecell[c]{$94.45\%$ \\ $\pm 0.373\%$}  & \makecell[c]{$0.162$ \\ $\pm 0.001$}  \\
        \midrule
        \makecell[l]{TinyResNet-KD \\ (with Residual-1) } & \makecell[c]{$93.48\%$ \\ $\pm 0.53\%$}  &\makecell[c]{$94.67\%$ \\ $\pm 0.373\%$} & \makecell[c]{$0.165$ \\ $\pm 0.001$} \\
        \makecell[l]{TinyResNet-KD \\ (with Residual-2) } & \makecell[c]{\textcolor{myblue}{$\mathbf{94.25}\%$} \\ \textcolor{myblue}{$\pm \mathbf{0.075\%}$}}  &\makecell[c]{\textcolor{myblue}{$\mathbf{95.56\%}$} \\ \textcolor{myblue}{$\pm \mathbf{0.572}\%$}} & \makecell[c]{\textcolor{myblue}{$\mathbf{0.144}$} \\ \textcolor{myblue}{$\pm \mathbf{0.002}$}} \\
        \makecell[l]{TinyResNet-KD \\ (with Residual-3) } & \makecell[c]{$94.06\%$ \\ $\pm 0.108\%$}  &\makecell[c]{$95.18\%$ \\ $\pm 0.62\%$} & \makecell[c]{$0.153$ \\ $\pm 0.002$}\\
        \bottomrule
        \vspace{0.05cm}
    \end{tabular}
\caption{Comparing the effects of adaptation module structure on the distillation process in a binary classification task. Unlike common approaches, using deep structures such as residual layers can further improve the distillation process and increase the performance of the student model. In this study, a 2-layer residual adaptation module has the best performance.}
\label{table:kd-class}
\end{table}

For the object detection task defined in Section~\ref{sec:exp}, we run ablation studies to understand the effect of the adaptation module in the knowledge distillation. Note that, the results presented here can be further improved when combined with other forms of knowledge distillation. However, the main goal here is to understand solely the effects of adaptation module structure. For this task, we consider adaptation from the backbone module of the edge model (\textsf{YOLOv5xs}) to the backbone of the cloud model (\textsf{YOLOv5l}), which is denoted by $\mathsf{B}\rightarrow\mathsf{B}$. For more information about this adaptation refer to Section~\ref{sec:struct}. As for the adaptation module, we first use a simple convolution layer that adapts the feature maps of the edge model to the cloud one by increasing their respective channel size. Then, we also try deeper architectures introduced by~\citet{wang2020cspnet} known as Bottleneck CSP. We also change the number of blocks of bottleneck modules to find the best module size. Table~\ref{table:kd-object} summarizes the results of this ablation study and comparing the performance of the learned models using knowledge distillation with a convolution layer and bottleneck modules, as well as the one without knowledge distillation. It can be inferred that by using deeper architectures such as bottleneck we can improve the distillation process in the form of hint layers. The model using $4$ blocks of bottleneck modules as the adaptation layer has the best performance in terms of $\mathsf{mAP}@0.5:0.95$. For this experiment, we set the confidence threshold to $0.001$.

\begin{table}[t]
    \centering
    \begin{tabular}{lcccc}
    \toprule
                    &  \multicolumn{4}{c}{Objectives}
        \\ \cmidrule(r){2-5}
         Model      & Precision & Recall  & $\mathsf{mAP}@0.5$ & $\mathsf{mAP}@0.5:0.95$
        \\
        \midrule
         \textsf{YOLOv5xs}  & \makecell[c]{$0.327$ \\ $\pm 0.008$}  &\makecell[c]{$0.41$ \\ $\pm 0.004$} & \makecell[c]{$0.344$ \\ $\pm 0.001$} & \makecell[c]{$0.153$ \\ $\pm 0.006$}
        \\
        \midrule
          \makecell[l]{\textsf{YOLOv5xs}-KD \\ (with Conv) }&\makecell[c]{\textcolor{myblue}{$0.336$} \\ \textcolor{myblue}{$\pm 0.001$}} & \makecell[c]{$0.437$ \\ $\pm 0.002$}  & \makecell[c]{$0.372$ \\ $\pm 0.001$}  & \makecell[c]{$0.167$ \\ $\pm 0.0003$} \\
        \midrule
        \makecell[l]{\textsf{YOLOv5xs}-KD \\ (with Bottleneck-2) } & \makecell[c]{$0.303$ \\ $\pm 0.003$}  &\makecell[c]{$0.486$ \\ $\pm 0.002$} & \makecell[c]{$0.397$ \\ $\pm 0.001$} & \makecell[c]{$0.185$ \\ $\pm 0.001$}\\
        \makecell[l]{\textsf{YOLOv5xs}-KD \\ (with Bottleneck-3) } & \makecell[c]{$0.304$ \\ $\pm 0.003$}  &\makecell[c]{\textcolor{myblue}{$\mathbf{0.489}$} \\ \textcolor{myblue}{$\pm \mathbf{0.004}$}} & \makecell[c]{\textcolor{myblue}{$\mathbf{0.402}$} \\ \textcolor{myblue}{$\pm \mathbf{0.004}$}} & \makecell[c]{$0.185$ \\ $\pm 0.001$}\\
        \makecell[l]{\textsf{YOLOv5xs}-KD \\ (with Bottleneck-4) } & \makecell[c]{$0.31$ \\ $\pm 0.005$}  &\makecell[c]{$0.483$ \\ $\pm 0.002$} & \makecell[c]{$0.4$ \\ $\pm 0.002$} & \makecell[c]{\textcolor{myblue}{$\mathbf{0.186}$} \\ \textcolor{myblue}{$\mathbf{\pm 0.001}$}}\\
        \bottomrule
        \vspace{0.05cm}
    \end{tabular}
\caption{Investigating the effects of adaptation module structure on the knowledge distillation process. Previous approaches mainly use simple structures such as a Conv Layer. We propose using deep architectures such as Bottleneck CSP modules~\cite{wang2020cspnet}. We also see the effect of changing the depth of Bottleneck modules on the distillation process by having $2$,$3$, and $4$ layers of Bottleneck.}
\label{table:kd-object}
\end{table}

\paragraph{Recall Rate Boosting} As it was mentioned, we can use the technique in Proposition~\ref{prop:pareto} in order to boost the recall rate while optimizing for accuracy as well. Solving the optimization in (\ref{eq:quad}) is straightforward using convex quadratic programming tools. At each iteration, we get the gradients for each objective (accuracy and recall), and solve the optimization (\ref{eq:quad}) to find the optimal weights for each objective's gradient direction. Using these weights, we generate the gradient direction as the weighted sum of objectives' gradients and update the parameters of the model with it. To show the effectiveness of this approach, we use it in our classification task on the edge model, in order to improve the recall rate of the \ecci model trained for this task. In Table~\ref{table:MOO-class}, we compare the results for this optimization as well as the edge model and the cloud model. It can be seen that using the technique for recall rate boosting the accuracy and recall rate of the edge model improves. Moreover, when combined with the cloud model to form an \ecci model there is a huge boost to the accuracy and recall rate of the \ecci model with recall rate boosting over the normal one. The \ecci model learned by this technique as well as knowledge distillation has the best performance for the \ecci model.

\begin{table}[t]
    \centering
    \begin{tabular}{lll}
    \toprule
                    &  \multicolumn{2}{c}{Objectives}
        \\ \cmidrule(r){2-3}
         Model      & Test Accuracy & Recall Rate 
        \\
        \midrule
          \makecell[l]{TinyResNet \\ (Normal Edge)}  & \makecell[c]{$77.32\%$ \\ $\pm 0.106\%$}  &\makecell[c]{$93.217\%$ \\ $\pm 0.359\%$} 
        \\
          \makecell[l]{TinyResNet-RB \\ (Recall Boosting) }&\makecell[c]{$77.56\%$ \\ $\pm 0.092\%$} & \makecell[c]{$94.583\%$ \\ $\pm 0.3\%$}  \\
        \midrule
        \makecell[l]{\ecci \\ (Normal Edge) } & \makecell[c]{$89.69\%$ \\ $\pm 0.106\%$}  &\makecell[c]{$94.117\%$ \\ $\pm 0.359\%$} \\
        \makecell[l]{\ecci \\ (Edge + RB) } & \makecell[c]{$90.30\%$ \\ $\pm 0.102\%$}  &\makecell[c]{$96.067\%$ \\ $\pm 0.341\%$} \\
        \makecell[l]{\ecci \\ (Edge + RB + KD) } & \makecell[c]{\textcolor{myblue}{$\mathbf{91.01}\%$} \\ \textcolor{myblue}{$\pm \mathbf{0.042}\%$}} &\makecell[c]{\textcolor{myblue}{$\mathbf{96.383}\%$} \\ \textcolor{myblue}{$\pm \mathbf{0.346}\%$}} \\
        \midrule
        \makecell[l]{ResNet18 \\ (Cloud) } & \makecell[c]{$91.830\%$ \\ $\pm 0.05\%$}  &\makecell[c]{$98.252\%$ \\ $\pm 0.125\%$} \\
        \bottomrule
        \vspace{2mm}
    \end{tabular}
\caption{Investigating the effectiveness of recall boosting technique on the classification task. It can improve both the accuracy and recall rate of the edge model. Besides, it can greatly improve the recall rate and accuracy of the \ecci model using this model as its edge. Combining this technique with knowledge distillation gives us the best performance for the \ecci model. (RB: Recall Boosting, KD: Knowledge Distillation)}
\label{table:MOO-class}
\end{table}

\begin{figure}
    \centering
    \begin{subfigure}{0.47\textwidth}
      \centering
      \includegraphics[width=\textwidth]{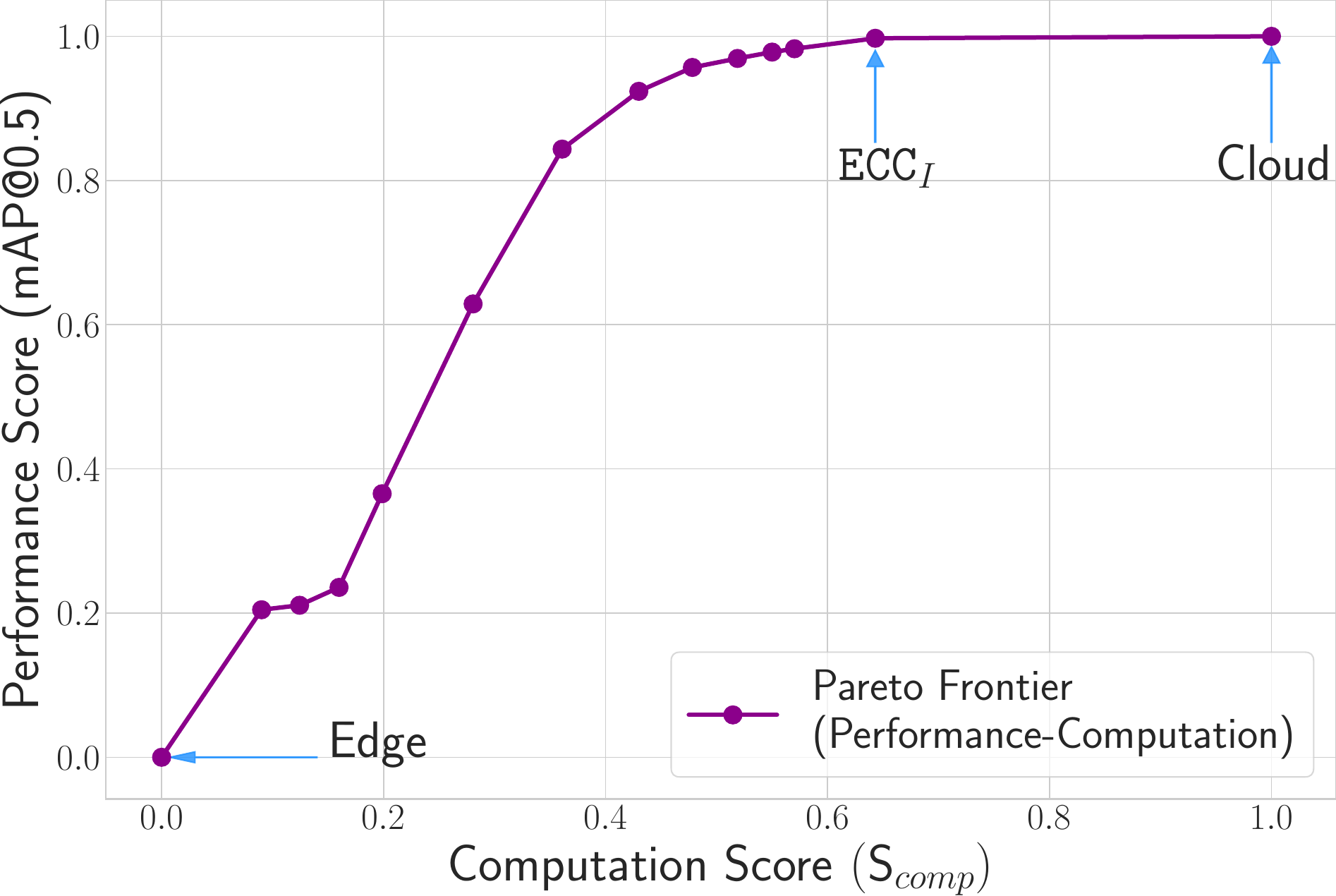}
      \caption{Performance vs. Computation}
    \end{subfigure}%
     \begin{subfigure}{0.47\textwidth}
      \centering
      \includegraphics[width=\textwidth]{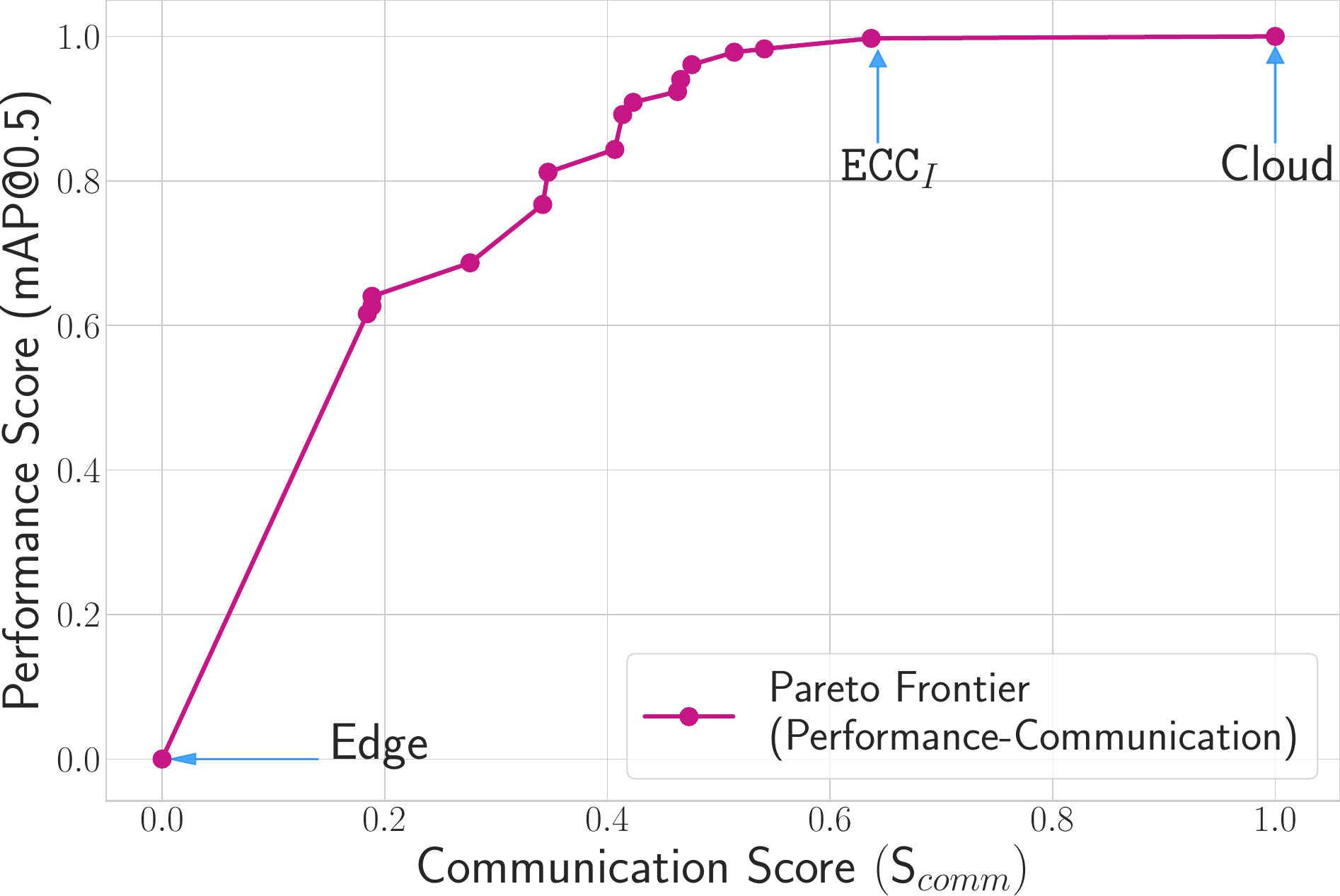}
      \caption{Performance vs. Communication}
    \end{subfigure}%
    \caption{The Pareto frontier curve of Performance vs. Computation and Performance vs. Communication extracted from the \ecc models learned for the defined object detection task.}
    \label{fig:pf}
\end{figure}

\paragraph{Pareto Frontier} As it was mentioned, we are seeking to find models that can fill the gap between two extremes of edge and cloud models in terms of computation, communication, and performance. Using \ecc framework we found different models with different levels of trade-offs between these objectives. If we only keep non-dominated models\footnote{Models with optimal trade-offs that no other model can dominate them in all objectives. For a bi-objective problem a model $\bm{w}_1$ is dominated if there is another model $\bm{w}_2$ that has better objective values in both objectives than $\bm{w}_1$.}, we can extract the Pareto frontier between those objectives. To illustrate that for the object detection tasks, we only keep the non-dominated models for trade-offs between performance vs. computation and performance vs. communication. Then connecting those points together, we can plot the Pareto frontier of these trade-offs as depicted in Figure~\ref{fig:pf}. The Pareto frontier curve will help us to decide which model to use based on the level of trade-off between models.

\subsection{Model Structures and Experimental Details}\label{sec:struct}
Empirical studies in this paper are run on p3.8xlarge instances of AWS with 4 GPUs for distributed training. The codes are developed using PyTorch and its distributed API.

\begin{figure}[t]
    \centering
    \includegraphics[width=0.6\textwidth]{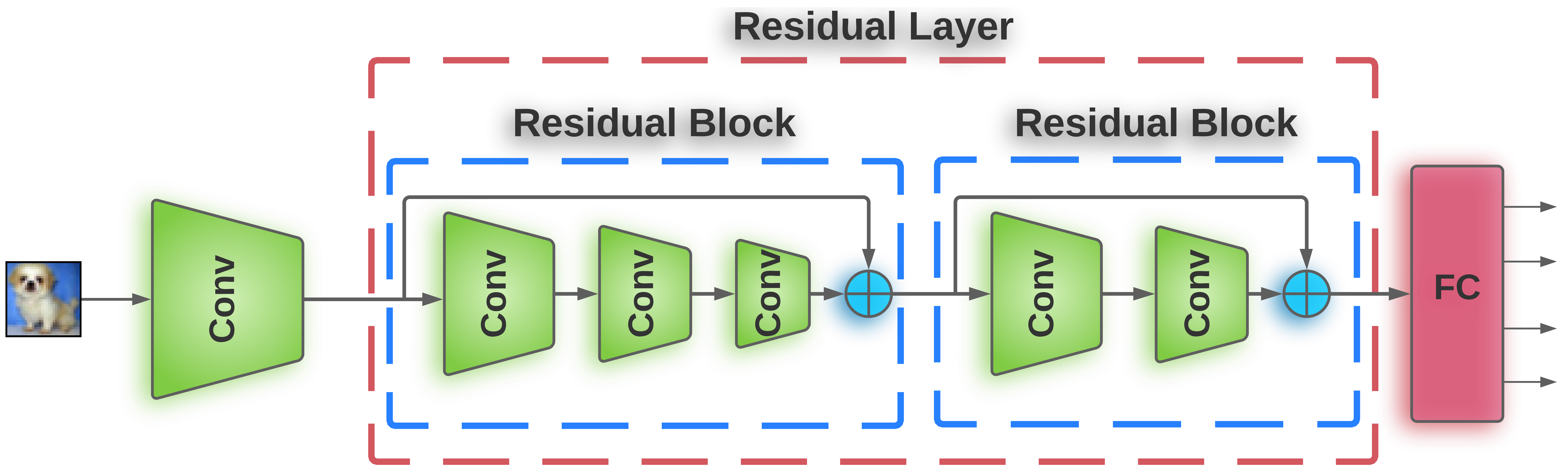}
    \caption{The structure of TinyResNet model, which is basically a ResNet5 with only one residual layer.}
    \label{fig:tinyresnet}
\end{figure}

\paragraph{Classification} For the classification task, we use a normal ResNet18 with $4$ residual layers as the cloud model and the teacher model. As for the edge model, we use a model we call TinyResNet, which only has one residual layer. Its structure is as shown in Figure~\ref{fig:tinyresnet}, where it has only one residual layer. As it was mentioned, for this task we use the CIFAR10 dataset and the goal is to classify between $7$ classes of $6$ animals and one non-animal class containing all other $4$ classes in the dataset.

\begin{figure}[t]
    \centering
    \includegraphics[width=\textwidth]{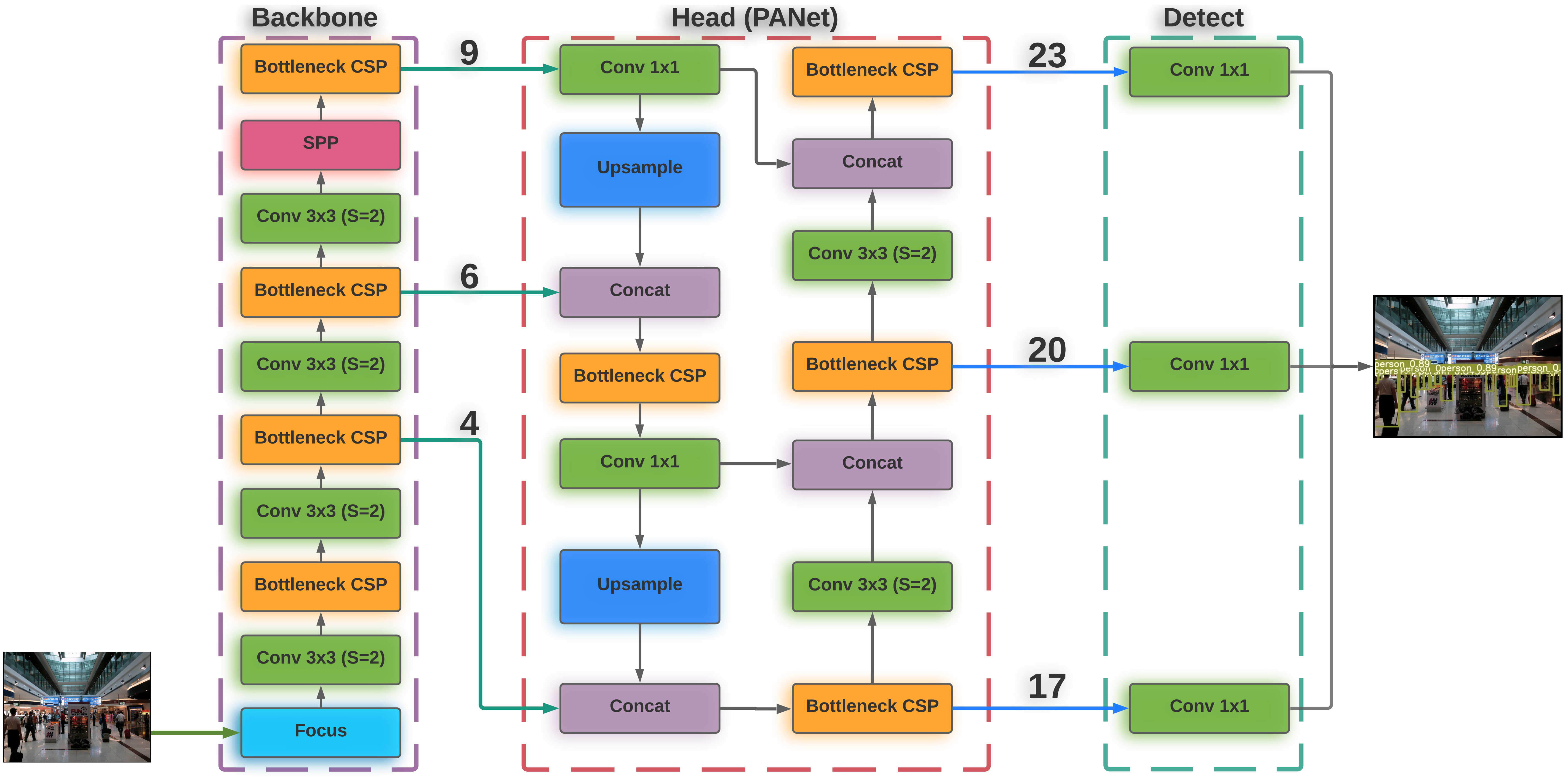}
    \caption{\textsf{YOLOv5} model structure.}
    \label{fig:yolov5}
\end{figure}

\paragraph{Object Detection} For the Object detection task we used model structures from \textsf{YOLOv5}\footnote{\url{https://github.com/ultralytics/yolov5}} as depicted in Figure~\ref{fig:yolov5}. For the teacher model or the cloud model, we use \textsf{YOLOv5l}, which is the large structure in the repository. For the edge model, we use \textsf{YOLOv5xs}, which we defined based on the main structure by reducing the depth and width multiplier of modules in the network. The model structure for both models is the same, but the width and depth of each module is different. Roughly the depth and width of the modules in the \textsf{YOLOv5l} is about $9$ times higher than its counterpart in the \textsf{YOLOv5xs}. In Section~\ref{sec:exp}, we use adaptation forms from the head of the edge model to the backbone of the cloud model ($\mathsf{H}\rightarrow\mathsf{B}$), from the head to the head ($\mathsf{H}\rightarrow\mathsf{H}$), and from the backbone to the backbone ($\mathsf{B}\rightarrow\mathsf{B}$). For the backbone adaptation, the feature maps are coming from layers $l\in\{4,6,9\}$, and for the head module adaptation, they are coming from layers $l\in\{17,20,23\}$ as depicted in Figure~\ref{fig:yolov5}. For an input size of $(3,512,512)$, the feature maps of the backbone coming from layers $4$, $6$, and $9$ of the \textsf{YOLOv5xs} are in the size of $(32,64,64)$, $(64,32,32)$, and $(128,16,16)$ respectively. These feature maps in a \textsf{YOLOv5l} model have the size of $(256,64,64)$, $(512,32,32)$, and $(1024,16,16)$ respectively. That means the spatial size is not changing from a model to another while the channel size is changing. The size of the feature maps is exactly the same for the head module outputs $17$, $20$, and $23$ as their counterparts in the backbone since the head has a pyramid feature map.

\section{Recall Rate Boosting}\label{app:recall}
As it was mentioned, in most cases the edge model is used as a filtration module for an \ecc model, to only pass samples with desired classes or objects for improved inference to the cloud. However, this requires a high recall rate for the edge model, not to miss any important sample, in addition to the main objective of the training for the edge model. To do so, we need to add another objective, which is the loss only on the positive samples\footnote{Positive samples are samples with classes other than normal in the classification and samples with at least one desired object in the object detection task.}. This loss will ensure to have a higher recall rate, but it might be at odds with the main training objective to some extent (decreasing precision rate), hence, we need to use multiobjective optimization approaches~\cite{miettinen2012nonlinear,cortes2020agnostic}, to ensure that both objectives are minimized simultaneously. To do so, we can reweight the objectives to converge to a point from the Pareto frontier of the problem. Instead of manually assigning the weights, we use the following proposition that ensures to find the best weights at every step to have a decreasing gradient direction for all objectives. 
\begin{proposition}[Pareto Descent Direction]
Consider a multiple objective problem with $p$ objectives of $\bm{\mathrm{h}}\left(\bm{w}\right) = \left[\mathrm{h}_1\left(\bm{w}\right), \mathrm{h}_2\left(\bm{w}\right), \ldots, \mathrm{h}_p\left(\bm{w}\right)\right]$, that ought to be minimized. Using the solution of the following quadratic optimization as the weights for each objective:
\begin{equation}\label{eq:quad}
    \bm{\alpha}^*(\bm{w}) \in \arg\underset{\bm{\alpha} \in \Delta_p}{\min} \left\|\sum_{i=1}^p \alpha_i\bm{\mathrm{g}}_i\left(\bm{w}\right)\right\|_2^2,
\end{equation}
where $\bm{\mathrm{g}}_i\left(\bm{w}\right) = \nabla_{\bm{w}}\mathrm{h}_i\left(w\right), i\in [p]$ and $\Delta_p$ is a $p$-dimensional simplex. We can show that the gradient of the weighted sum of objectives, using the solution of (\ref{eq:quad}) as the weights, is either zero or a descent direction for all the objectives. Meaning, we have:
\begin{equation}\label{eq:paretcond}
    - \left\langle \sum_{i\in [p]} \alpha^*_i \bm{\mathrm{g}}_i\left(\bm{w}\right), \bm{\mathrm{g}}_j(\bm{w}) \right\rangle\leq 0, \;\; \forall j\in\{1,\ldots,p\}\;.
\end{equation}
\label{prop:pareto}
\end{proposition}
This proposition will ensure that at every step of the gradient descent we will not increment any of the objectives until reaching a point from the Pareto frontier. The proof of this proposition is deferred to Appendix~\ref{app:pareto}.

section{Proof of Proposition~\ref{prop:pareto}}\label{app:pareto}
\begin{proof}
To prove this proposition we use contradiction to show that if the condition on (\ref{eq:paretcond}) was not correct, we reach a condition that is always wrong. First, for the model parameters in the $t$-th iteration of gradient descent step, $\bm{w}_t$, we define the feasible set of solutions for the optimization in (\ref{eq:quad}) as follows:
\begin{equation}
    {\Phi}(\bm{w}_t) = \left\{ \sum_{i=1}^{m} \alpha_i \bm{\mathrm{g}}_i\left(\bm{w}_t\right) \;|\; \alpha_i \geq 0 \;\;\forall i \in \{1,\ldots,m \}, \sum_{i=1}^m \alpha_i = 1  \right\}\;.
\end{equation}
Hence, based on the solution of (\ref{eq:quad}) in $\bm{w}_t$, the aggregated gradient for updating the parameter will be defined as:
\begin{equation}
    \bar{\bm{\mathrm{g}}}\left(\bm{w}_t\right) = \sum_{i=1}^m \alpha_i^*\left(\bm{w}_t\right) \bm{\mathrm{g}}_i\left(\bm{w}_t\right)
    \label{eq:descent_dir}\;.
\end{equation}
Thus, we consider that, if (\ref{eq:paretcond}) was wrong, then there would be at least one $\bm{\mathrm{g}}_i\left(\bm{w}_t\right) \in {\Phi}(\bm{w}_t)$, where $\langle\bar{\bm{\mathrm{g}}}\left(\bm{w}_t\right), \bm{\mathrm{g}}_i\left(\bm{w}\right)\rangle < 0$. Then, considering the following optimization problem,
\begin{equation}\label{eq:proof-opt}
    \underset{\eta \in [0,1]}{\min} \left\|(1-\eta) \bm{\mathrm{g}}_i\left(\bm{w}_t\right) + \eta \bar{\bm{\mathrm{g}}}\left(\bm{w}_t\right) \right\|_2^2\;,
\end{equation}
it can be verified that $(1-\eta) \bm{\mathrm{g}}_i\left(\bm{w}_t\right) + \eta \bar{\bm{\mathrm{g}}}\left(\bm{w}_t\right) \in \Phi(\bm{w}_t)$. Hence, its solution is a solution for (\ref{eq:quad}) as well. Also, from (\ref{eq:descent_dir}), it is defined that $\|\bar{\bm{\mathrm{g}}}^{(t)}\|_2^2$ is the solution to the (\ref{eq:quad}). Thus, it shows that the function in (\ref{eq:proof-opt}) is monotonically decreasing in $\eta$, and the minimum happens at $\eta^*=1$. The first order condition in the optimization problem of (\ref{eq:proof-opt}) can be written as:
\begin{align}\nonumber
    2\Big( \bar{\bm{\mathrm{g}}}\left(\bm{w}_t\right) - \bm{\mathrm{g}}_i\left(\bm{w}_t\right)\Big)^\top \Big( \bm{\mathrm{g}}_i\left(\bm{w}_t\right) + \eta\big(\bar{\bm{\mathrm{g}}}\left(\bm{w}_t\right) - \bm{\mathrm{g}}_i\left(\bm{w}_t\right)\big)\Big) &\leq 0 \\\nonumber
    \Big(\bar{\bm{\mathrm{g}}}\left(\bm{w}_t\right)\Big)^\top\Big( \bar{\bm{\mathrm{g}}}\left(\bm{w}_t\right) - \bm{\mathrm{g}}_i\left(\bm{w}_t\right)\Big) & \leq 0 \\\label{eq:proof-cont}
    \Big\|\bar{\bm{\mathrm{g}}}\left(\bm{w}_t\right)\Big\|_2^2 & \leq \Big(\bar{\bm{\mathrm{g}}}\left(\bm{w}_t\right)\Big)^\top\bm{\mathrm{g}}_i\left(\bm{w}_t\right)\;,\end{align}
where we replace $\eta$ with its optimum value $\eta^*=1$ in the second line. It can be inferred that inequality in (\ref{eq:proof-cont}) contradicts the assumption we had on $\langle\bar{\bm{\mathrm{g}}}\left(\bm{w}_t\right), \bm{\mathrm{g}}_i\left(\bm{w}\right)\rangle < 0$. Thus, we can conclude that $\bar{\bm{\mathrm{g}}}\left(\bm{w}_t\right)$ is a descent direction for all objectives, which is $-\langle\bar{\bm{\mathrm{g}}}\left(\bm{w}_t\right),\bm{\mathrm{g}}_i\left(\bm{w}_t\right)\rangle < 0$ for every $1 \leq i \leq p$.
\end{proof}

{\small
\begin{@fileswfalse}
\bibliography{ref}
\end{@fileswfalse}
}

%% file: 2-related.tex
\section{Additional Related Work}\label{sec:related}
\paragraph{Compression Methods} Compression methods have been widely used in many domains to minimize the memory footprint and computational power required for machine learning models to run their inference. The main technique that is vastly used in different applications is quantization, where the goal is to quantize the weights of the model to a lower bit precision, in order to benefit from faster computation and lower memory~\cite{li2017training,banner2018post,choi2019accurate,hubara2017quantized,Zhou2016}. This procedure will degrade the performance of the model since the learned weights are quantized and might not be optimal for the task at hand. Hence, different methods try to minimize the degradation effects of quantization by post-training mechanisms such as fine-tuning the quantized model~\cite{banner2018post}, or quantization-aware training~\cite{choi2019accurate,Zhou2016}. Another form of compression is stemming from the idea that the models are mostly over-parameterized, and hence, the parameter space is highly sparse. Hence, by pruning sparse weights we can reduce the model size and computation power required for the inference~\cite{han2015learning,li2016pruning, Zhu2017,Han2015}. The other major approach for compression is Knowledge Distillation, which will be described in the next part. These approaches can compress the model to a certain degree, however, there is a lower bound on their compression rate. Meaning, starting from a large model, these compression methods can compress it to some extent before heavily sacrificing its performance. 

\paragraph{Knowledge Distillation}
Knowledge distillation was first introduced by~\citet{hinton2015distilling} as a new method for compression. This method, instead of starting with a big model and compress it, tries to distill the knowledge from a bigger model, called the teacher model, to a smaller model, named the student model. It is shown that this technique can boost the performance of the student model to certain degrees. Theoretically, the student model can be as small as possible, however, by reducing the size of the student model, the gap between the performance of the student and teacher models increases. Although knowledge distillation approaches can boost the performance of the student model, they certainly cannot fill this gap. Hence, practically, these approaches cannot have a reasonable impact on very small models suitable for low-powered edge devices. Nonetheless, the ideas from knowledge distillation are inspirational motives of this work to fill this gap using a collaboration between edge and cloud models. \citet{hinton2015distilling} first introduced this approach for classification models by transferring knowledge from the classification output of the teacher model to its counterpart in the student model. \citet{romero2014fitnets} introduced FitNets, where the distillation can happen between any two layers of neural network using matching modules. Since then, there are so many different forms of distillation introduced by various studies~\cite{tung2019similarity,park2019relational,wang2018dataset,radosavovic2018data,mishra2017apprentice}. In addition to classification, there are approaches introduced for other tasks as well. More specifically, \citet{chen2017learning} introduced using classification and hint layers for distilling knowledge in object detection models. \citet{wang2018dataset} introduced using a mask generated by the mistakes of the student models to have more attention during knowledge distillation. 

\paragraph{Feature Adaptation} Domain adaptation methods have been using similar techniques to FitNets~\cite{romero2014fitnets} to adapt feature layers from the target domain to the source domain, in order to improve the performance of the model in the source domain~\cite{long2016unsupervised, sankaranarayanan2018generate,li2020model}. The adaptation happens mostly between same model structures, however, there have been some recent studies of domain adaptation in the edge models~\cite{jiang2020resource,yang2020mobileda}. Nonetheless, in all these frameworks the goal is learn a standalone model for the target domain, and hence, the feature adaptation parts are not used during inference. This is unlike our approach, where the domain is the same, but the goal is to learn a model from both edge and cloud using feature adaptation methods and use them during inference as well.\cite{zeng2020coedge,fang2019teamnet,xu2020collaborative,huang2019deepar}